\documentclass[11pt]{article}
\usepackage{graphicx}
\usepackage{amssymb}
\usepackage{dcolumn}
\usepackage{bm}

\title{Partially coherent matter wave and its evolution}

\author{\begin{tabular}{c}Jun Chen, Qiang Lin \\
Department of Physics, Zhejiang University, Hangzhou 310027, China
\end{tabular}}
\date{}

\begin{document}
\maketitle

\begin{abstract}
The evolution and propagation of a partially coherent matter wave
(PCMW) is investigated theoretically by the correlation function
method. The ABCD matrix formalism previously used for a fully
coherent matter wave is extended to the PCMW domain. A new ABCD law
is derived, using a tensor method to describe the evolution of a
PCMW. An analytical solution of the first-order correlation function
is obtained that makes the propagation and evolution of a PCMW very
simple and clear. As an example, the propagation of a PCMW in a
gravitational field is calculated numerically.
\end{abstract}

\section{Introduction}
\label{} Matter-wave optics is a new and fascinating branch of
research benefiting from the rapid progress of techniques for laser
cooling and trapping of neutral atoms and molecules. The achievement
of Bose-Einstein condensation (BEC) in dilute atomic and molecular
gases enables the study of matter waves from incoherent to coherent
domains \cite{coherMW}. In recent years, coherent matter waves have
been studied extensively, such as the formation of vortices in BEC,
linear and nonlinear propagation of atomic laser beams
\cite{nonlinearMW,BEClaser,BEC}, and other phenomena. The matter
wave has unique applications, as in atomic interferometry
\cite{interferometry}, atomic holography
\cite{holography1,holography2}, and so on.

According to existing theory, pure BEC exists only at zero
temperature. However, zero temperature is not reachable
experimentally due to the limits of laser cooling \cite{cooling}.
All the BEC and atomic lasers employed in experiments have
distributions of momenta \cite{BEC}. Therefore, matter waves, in the
strict sense, are not completely coherent. All the cold atom gases
produced by laser cooling at finite temperature are partially
coherent matter waves (PCMW), whose properties during propagation
and evolution are important issues and play a crucial role in the
applications of matter waves. Nevertheless, they not been
investigated sufficiently.

It is well known that a coherent matter wave can be described by a
single wave function whose evolution is governed by the Gross
Pitaevskii (GP) equation \cite{BEC}. On the other hand, a PCMW can't
be described by a single-particle wave function, but requires a much
more complicated treatment. One method is the finite temperature BEC
theory, which separates the matter wave field operator into two
terms; one term describes the condensate part, and the other term
deals with the noncondensate part. However, the equations based on
this method are complicated and can't be solved analytically.
Additional approximations are necessary to solve the equations. For
example, in the Hartree-Fock-Bogliubov (HFB) approximation
\cite{HFB}, the condensate wave function satisfies the GP equation,
and the noncondensate operator is described by an equation that
follows from subtracting the GP equation from the total field
operator's Heisenberg equation. In the Hartree-Fock (HF) and Popov
approximation \cite{HFP}, the condensate wave function is defined by
the modified GP equation, whereby the \textquotedblleft anomalous"
density is neglected and the noncondensate part consists of
thermally excited atoms as described in terms of a semiclassical
phase space distribution function. In an extended HFB theory, the
collisionless noncondensate dynamics is included within second-order
perturbation schemes, and the expressions for damping rates and
frequency shifts of low-energy modes are derived \cite{exHFB}.

Another approach to describe a PCMW is the correlation function,
which provides quantitative information about the coherence and the
intensity of matter waves. The first-order correlation function
characterizes local fluctuations of the phase of the complex matter
wave field amplitude, and is related to the contrast achievable in
an interference experiment. The second-order correlation function is
related to fluctuations of the modulus of the complex matter wave
field amplitude, and expresses the tendency of atoms either to
cluster or to remain spatially separated \cite{glauber}. Meystre et
al. derived the Van-Citter-Zernike theorem of PCMW, which disregards
the interactions among the particles. They analyzed the propagation
of the matter wave from an incoherent matter wave source \cite
{meystre94}, and discussed the detection of the correlation function
\cite{meystre98}. This method is analogous to that used in
traditional optics \cite{optical coherence,lin cai,lin w}. In this
paper, we focus on the propagation and evolution of PCMW by applying
the correlation function method, and disregard the interactions
among the particles, although it is not necessary. The ABCD matrix
formalism \cite{borde,borde2}, which previously had only been
applicable for the wave function of a single particle, is extended
to the PCMW. A generalized ABCD law to describe the evolution of the
PCMW is derived by a tensor method. An analytical solution of the
first-order correlation function after evolution is obtained, which
makes the evolution problem of the PCMW very simple and clear. The
results are useful in the analysis of the space-time coherence,
spatial distribution, propagation and imaging properties of the
PCMW.

This paper is organized as follows: In Sec.\ref{bordABCD}, we
introduce
briefly the ABCD matrix formalism for a single-particle wavepacket. In Sec.%
\ref{tensorABCD}, the ABCD matrix formalism is extended by a tensor
method to describe the PCMW. In Sec.\ref{example}, the evolution
properties of an ultracold atomic sample released in a gravitational
field are calculated numerically and discussed. A summary is given
in the Conclusion.

\section{The ABCD matrix formalism}
\label{} \label{bordABCD} First, let us briefly describe the ABCD
matrix formalism for a matter wave, introduced by Ch. J. Bord\'{e}
based on the time-dependent Schr\"{o}dinger equation \cite{borde},
which can be used to calculate the evolution of a single atom
wavepacket $|\psi (t)>$:
\begin{equation}
i\hbar \frac{\partial |\psi (t)>}{\partial t}=\hat{H}|\psi (t)>.
\end{equation}%
This equation can be solved in terms of the quantum
propagator\cite{borde}, which can be obtained by the use of a
shortcut through the classical limit, and a well-known result of Van
Vleck that makes the connection with quantum mechanics.

The Hamiltonian of a single atom with a mass $m$ in a
gravito-inertial field in the classical limit is
\begin{eqnarray}
H&=& \mathbf{p}^{T}\mathbf{\bm\beta }\mathbf{p}/(2m)-\bm\Omega ^{T}\bm{L}-m%
\mathbf{g}^{T}\mathbf{q}  \label{H1} \nonumber\\
&& -m\mathbf{q}^{T}\bm\gamma \mathbf{q}/2+V(t),
\end{eqnarray}%
where $V(t)$ is the external field, $\mathbf{q}^{T}=(x,y,z)$ is the
position
vector, $\mathbf{p}^{T}=(p_{x},p_{y},p_{z})$ is the momentum vector, and $%
\bm{L}$ is the angular momentum vector of the atom. The
gravitational wave
and gravitational gradient are represented by the tensors $\bm\beta $ and $%
\bm\gamma $. The quantity $\bm\Omega $ is the angular velocity
vector of the Earth's rotation, and $\mathbf{g}$ is the
gravitational acceleration vector. The Earth's gravito-inertial
field is characterized by the above four parameters: $\bm\beta
,\bm\gamma ,\bm\Omega ,\mathbf{g}$. The superscript $T$ means the
transpose. Some judiciously chosen transformations and choices
lead to the following ABCD formalism for the evolution of the variables $%
\mathbf{q}$ and $\mathbf{p}$ from the initial time $t_{a}$ to the
final time $t_{b}$\cite{borde}:
\begin{equation}\label{abcdq}
\left(
\begin{array}{c}
\mathbf{q}_{b} \\
\mathbf{p}_{b}/m%
\end{array}%
\right) =\left(
\begin{array}{c}
\bm U\bm\xi  \\
\bm n^{-1}\bm U\dot{\bm\xi }%
\end{array}%
\right) +\left(
\begin{array}{ll}
\mathbf{A} & \mathbf{B} \\
\mathbf{C} & \mathbf{D}%
\end{array}%
\right) \left(
\begin{array}{c}
\mathbf{q}_{a} \\
\mathbf{p}_{a}/m%
\end{array}%
\right) ,
\end{equation}%
where $\bm n^{-1}$ is the matrix corresponding to the gravitational
wave
tensor $\bm\beta $. Were we to neglect the gravitational wave effect, then $%
\bm n^{-1}$ would be a unit matrix. The matrix $\bm U$ corresponds
to the rotation term $-\mathbf{\Omega }^{T}\bm L$ in Eq.(\ref{H1}).
If there is no
angular momentum of the atom, $\bm U$ is also a unit matrix. The vector $\bm%
\xi $ is the displacement produced by gravity, which describes the
classical trajectory of a non-rotating Hamiltonian with the initial
conditions $\bm\xi
(t_{a})=0$ and $\dot{\bm\xi }(t_{a})=0$. The $3\times 3$ matrices $\mathbf{%
A,B,C,D}$ and the vector $\bm\xi $ can be determined by solving the
Hamilton-Jacobi equation\cite{borde}.

From the above expressions, by neglecting the gravitational wave
effect and the rotation, the classical action can be written as
\begin{eqnarray}
&S&(\mathbf{q}_{b},t_{b},\mathbf{q}_{a},t_{a}) \label{action} \nonumber\\
&=& m\dot{\bm\xi }\cdot (\mathbf{q}_{b}-\bm\xi
)+\int_{t_{a}}^{t_{b}}L_{1}(t_{1})dt_{1}-\int_{t_{a}}^{t_{b}}V(t_{2})dt_{2}
\nonumber\\
&&+\frac{m}{2}[(\mathbf{q}_{b}^{T}-\bm\xi ^{T})\mathbf{{DB}^{-1}(\mathbf{q}%
_{b}-\bm\xi )-2(\mathbf{q}_{b}^{T}-\bm\xi
^{T})(B^{-1})^{T}\mathbf{q}_{a}}
\nonumber\\
&&+\mathbf{q}_{a}^{T}\mathbf{{B^{-1}A}{q}_{a}],}
\end{eqnarray}%
where $L_{1}(t)$ is a partial Lagrangian given by $L_{1}(t)=m(|\dot{\bm\xi }%
|^{2}+\bm\xi ^{T}\bm\gamma \bm\xi +2\mathbf{g}^{T}\bm\xi )/2$.
Knowing the classical action allows the determination of the quantum
propagator according to Van Vleck's formula\cite{vleck}:
\begin{equation}\label{K}
K(\mathbf{q}_{b},t_{b},\mathbf{q}_{a},t_{a})=\left(
\frac{m}{ih}\right) ^{3/2}\left\vert \det \mathbf{B}\right\vert
^{-1/2}\exp (iS_{b,a}/\hbar ),
\end{equation}%
where $h$ is the Plank constant and $\hbar =h/(2\pi )$. A complete
set of solutions of the Schr\"{o}dinger equation can be derived from
$K$. The lowest order Gaussian wavepacket at initial time $t_{a}$ is
given by
\begin{eqnarray}
\psi (\mathbf{q},t_{a})=&&\psi
_{wp}(\mathbf{q},t_{a};\mathbf{q}_{a},\bm{v}_{a},\mathbf{X_{a},Y_{a})}  \nonumber \\
=&& \frac{1}{\sqrt{\det \mathbf{X_{a}}}}\exp
[\frac{im}{2\hbar}(\mathbf{q}^{T}-\mathbf{q}_{a}^{T})\mathbf{Y}_{a}
\mathbf{X}_{a}^{-1}(\mathbf{q}-\mathbf{q}_{a})] \nonumber\\
&&\times \exp
[\frac{im}{\hbar}\bm{v}_{a}\cdot(\mathbf{q}-\mathbf{q}_{a})],
\end{eqnarray}
which is centered at position $\mathbf{q}_{a}$, has an average velocity $%
\bm{v}_{a}=\mathbf{p}_{a}/m,$ and the complex width parameters $\mathbf{X_{a}%
}$, $\mathbf{Y_{a}}$. Its evolution
\begin{eqnarray}
\psi (\mathbf{q},t_{b}) =&\int K(\mathbf{q},t_{b},\mathbf{q^{\prime }}%
,t_{a})\psi _{wp}(\mathbf{q^{\prime }},t_{a};\mathbf{q}_{a},\bm{v}_{a},%
\mathbf{X_{a},Y_{a})d\mathbf{q^{\prime }}}  \nonumber \\
=& \exp [\frac{iS(\mathbf{q}_{b},t_{b},\mathbf{q}_{a},t_{a})}{\hbar
}]\psi
_{wp}(\mathbf{q},t_{a};\mathbf{q}_{b},\bm{v}_{b},\mathbf{X_{b},Y_{b})}
\end{eqnarray}%
is governed by the ABCD matrix formalism for $\mathbf{q}$ and
$\mathbf{p}$ in Eq.(\ref{abcdq}) and for $\mathbf{X}$ and
$\mathbf{Y}$:
\begin{eqnarray}
\mathbf{X_{b}}& \mathbf{=AX_{a}+BY_{a}},  \\
\mathbf{Y_{b}}& \mathbf{=CX_{a}+DY_{a}.}
\end{eqnarray}

From the ABCD matrix formalism, the problem of the evolution
of a single particle wavepacket can be solved by calculating the matrices $%
\mathbf{A,B,C}$ and $\mathbf{D}$, from which the classical action
and the classical equations of motion can be obtained.

As a many-particle system, such as the PCMW, can't be characterized
by a single-particle wavepacket, the theory must be extended. A
convenient approach to this problem is to carry out a second
quantization, which will be done in the following section.

\section{Tensor ABCD law for the PCMW}
\label{} \label{tensorABCD}

 In second quantization, atoms can be described by a quantum field
operator $\hat{\psi}(\mathbf{q},t)$ that satisfies the commutation
relations
\begin{eqnarray}
\lbrack \hat{\psi}(\mathbf{q},t),\hat{\psi}^{\dag
}(\mathbf{q}^{\prime
},t)]_{\pm }& =&\delta (\mathbf{q}-\mathbf{q}^{\prime }),  \\
\lbrack \hat{\psi}(\mathbf{q},t),\hat{\psi}(\mathbf{q}^{\prime
},t)]_{\pm }& =&0,
\end{eqnarray}%
where $\hat{\psi}^{\dag }$ is the conjugate operator of
$\hat{\psi}$. The expression $[\cdots ]_{-}$ is a commutator for
bosons, while $[\cdots ]_{+}$
is an anticommutator for fermions. The field operator $\hat{\psi}(\mathbf{q}%
,t)$ is interpreted as annihilating a particle at $(\mathbf{q},t),$ and $%
\hat{\psi}^{\dag }(\mathbf{q},t)$ represents the creation of a particle at $(%
\mathbf{q},t)$. The physical properties of a many-atom gas can be
expressed in terms of correlation functions that are expectation
values of the field operators, such as the first-order correlation
function
\begin{equation}\label{Gamma0}
\Gamma (\mathbf{q}_{1},t_{1},\mathbf{q}_{2},t_{2})=\langle \hat{\psi}%
^{\dagger
}(\mathbf{q_{1}},t_{1})\hat{\psi}(\mathbf{q_{2}},t_{2})\rangle ,
\end{equation}%
and its higher orders: $\langle \hat{\psi}^{\dagger }(\mathbf{q_{1}},t_{1})%
\hat{\psi}^{\dagger }(\mathbf{q_{2}},t_{2})\cdots \hat{\psi}(\mathbf{%
q_{2}^{\prime }},t_{2}^{\prime })\hat{\psi}(\mathbf{q_{1}^{\prime }}%
,t_{1}^{\prime })\rangle .$ Here we shall restrict ourselves to the
first-order correlation function of bosonic particles in the form of Eq.(\ref%
{Gamma0}). Its diagonal element $\Gamma (\mathbf{q},t,\mathbf{q},t)$
is the atom number density at position $\mathbf{q}$ and time $t$.

For a completely coherent matter wave, the first-order correlation
function can be factorized into the form\cite{glauber}
\begin{equation}\label{factorize}
\langle \hat{\psi}^{\dagger }(\mathbf{q_{1}},t_{1})\hat{\psi}(\mathbf{q_{2}}%
,t_{2})\rangle =\psi ^{\ast }(\mathbf{q}_{1},t_{1})\psi (\mathbf{q}%
_{2},t_{2}),
\end{equation}%
with
\begin{equation}
\psi (\mathbf{q},t)=\sqrt{N}\langle \mathbf{q}|\phi (t)\rangle ,
\end{equation}%
where atoms occupy the same one-particle state $|\phi (t)\rangle $,
and $N$ is the total atom number. The degree of coherence
\begin{equation}
g(\mathbf{q}_{1},t_{1},\mathbf{q}_{2},t_{2})=\frac{\Gamma (\mathbf{q}%
_{1},t_{1},\mathbf{q}_{2},t_{2})}{\sqrt{\Gamma (\mathbf{q}_{1},t_{1},\mathbf{%
q}_{1},t_{1})}\sqrt{\Gamma
(\mathbf{q}_{2},t_{2},\mathbf{q}_{2},t_{2})}}
\end{equation}%
is unity for a perfectly coherent matter wave such as the pure BEC,
which contains millions of atoms associated with a single-particle
wave function.

The ABCD matrix formalism in Sec.\ref{bordABCD} deals with the
evolution of a one-particle wavepacket and is appropriate to
describe a coherent matter wave disregarding interactions among
atoms \cite{coq,coq06,atom phase,gravimeter}. If the coherence of
the matter wave is not perfect, that is, if
\begin{equation}
0<g(\mathbf{q}_{1},t_{1},\mathbf{q}_{2},t_{2})<1,
\end{equation}%
it is not possible to factorize the first-order correlation function
into the form of Eq.(\ref{factorize}). That means that the matter
wave field is only partially coherent and some randomness exists.
The field can't just be represented by the single-particle wave
function, and second quantized field theory has to be applied.

We concentrate on the propagation and evolution of non-interacting
matter waves from a PCMW source. Neglecting the gravitational-wave
effects and rotational effects, the single-particle Hamiltonian of
Eq.(\ref{H1}) turns out to be
\begin{equation}
H=\mathbf{p}^{T}\mathbf{p}/(2m)-m\mathbf{g}^{T}\mathbf{q}-m\mathbf{q}^{T}\bm%
\gamma \mathbf{q}/2.
\end{equation}%
Changing this single-particle Hamiltonian to the second quantized
Hamiltonian, the Heisenberg equation of motion for the field operator $\hat{%
\psi}(\mathbf{q},t)$ can be found\cite{atomoptics}:
\begin{equation}\label{heisenberg}
i\hbar \frac{\partial }{\partial {t}}\hat{\psi}(\mathbf{q},t)=\left( -\frac{%
\hbar ^{2}\nabla ^{2}}{2m}-m\mathbf{g}^{T}\mathbf{q}-\frac{m}{2}\mathbf{q}%
^{T}\bm\gamma \mathbf{q}\right) \hat{\psi}(\mathbf{q},t).
\end{equation}%
It has the same form as the Schr\"{o}dinger equation for the
single-particle wave function, so solution techniques are similar.
The propagation of a matter wave field is described by the quantum
mechanical propagator
\begin{equation}\label{path-integral}
\hat{\psi}(\mathbf{q},t)=\int K(\mathbf{q},t,\mathbf{q_{0}},t_{0})\hat{\psi}(%
\mathbf{q_{0}},t_{0})d\mathbf{q_{0}}.
\end{equation}%
The action Eq.(\ref{action}) can be rewritten in tensor form as
\begin{eqnarray}
&S& (\mathbf{q},t,\mathbf{q}_{0},t_{0})   \label{action2} \nonumber\\
&& =m\dot{\bm\xi }\cdot (\mathbf{q}-\bm\xi
)+\int_{t_{0}}^{t}L_{1}(t_{1})dt_{1}-\int_{t_{0}}^{t}V(t_{2})dt_{2}
\nonumber\\
\ && +\frac{m}{2}\left(
\begin{array}{c}
\mathbf{q_{0}} \nonumber\\
\mathbf{q}-\bm\xi
\end{array}%
\right) ^{T}\left(
\begin{array}{ll}
\mathbf{B^{-1}A} & \mathbf{-B^{-1}} \nonumber\\
\mathbf{C-DB^{-1}A} & \mathbf{DB^{-1}}%
\end{array}%
\right) \left(
\begin{array}{c}
\mathbf{q_{0}}\nonumber \\
\mathbf{q}-\bm\xi
\end{array}%
\right)  \nonumber \\
&
\end{eqnarray}%
From Eqs.(\ref{K}), (\ref{Gamma0})and (\ref{path-integral}), we find
that the general propagation formula for the first-order correlation
function is
\begin{equation}\label{Gamma}
\Gamma (\mathbf{r},t)=-\left( \frac{m}{ih}\right) ^{3}\frac{1}{|\det \mathbf{%
B}|}\int \Gamma (\mathbf{r_{0}},t_{0})\exp \left( -i\frac{\Delta S}{\hbar }%
\right) d\mathbf{{r}_{0}},
\end{equation}%
where $\mathbf{r}$ and $\mathbf{r}_{0}$ are position tensors given by $%
\mathbf{r}^{T}=(\mathbf{q}_{1}^{T},\mathbf{q}_{2}^{T})$, and $\mathbf{r}%
_{0}^{T}=(\mathbf{q}_{01}^{T},\mathbf{q}_{02}^{T})$; $\Delta
S=(S_{1}-S_{2})$ is the action difference, where $S_{j}(j=1,2)$ is
the action from the incident point $\mathbf{q}_{0j}$ to the output
point $\mathbf{q}_{j}$, and
can be obtained from Eq.(\ref{action2}) by replacing $\mathbf{q}$ with $%
\mathbf{q}_{j}$, $\mathbf{q}_{0}$ with $\mathbf{q}_{0j}$. The action
difference is

\begin{equation}\label{deltaS}
\Delta S=m\dot{\bm\xi }\cdot (\mathbf{q}_{1}-\mathbf{q}_{2})+\frac{m}{2}%
\left(
\begin{array}{c}
\mathbf{r_{0}} \\
\mathbf{r-\bar{\bm\xi }}%
\end{array}%
\right) ^{T}\mathbf{V}\left(
\begin{array}{c}
\mathbf{r_{0}} \\
\mathbf{r-\bar{\bm\xi }}%
\end{array}%
\right) ,
\end{equation}%
where $\bar{\bm\xi }^{T}=(\bm\xi ^{T},\bm\xi ^{T})$,

\begin{equation}\label{V}
\mathbf{V}=\left[
\begin{array}{cc}
\mathbf{\bar{B}^{-1}\bar{A}} & \mathbf{-\bar{B}} \\
\mathbf{\bar{C}-\bar{D}\bar{B}^{-1}\bar{A}} & \mathbf{\bar{D}\bar{B}^{-1}}%
\end{array}%
\right] ,
\end{equation}%
and
\begin{eqnarray}
\mathbf{\bar{A}}& =\left[
\begin{array}{cc}
\mathbf{A} & \bm0 \\
\bm0 & \mathbf{A}%
\end{array}%
\right] ,\ \ \ \ \mathbf{\bar{B}}=\left[
\begin{array}{cc}
\mathbf{B} & \bm0 \\
\bm0 & -\mathbf{B}%
\end{array}%
\right] ,   \\
\mathbf{\bar{C}}& =\left[
\begin{array}{cc}
\mathbf{C} & \bm0 \\
\bm0 & -\mathbf{C}%
\end{array}%
\right] ,\ \ \ \ \mathbf{\bar{D}}=\left[
\begin{array}{cc}
\mathbf{D} & \bm0 \\
\bm0 & \mathbf{D}%
\end{array}%
\right] .
\end{eqnarray}%
Since $\Delta S$ is a scalar, we get the relations
\begin{eqnarray}
&&(\mathbf{\bar{B}^{-1}\bar{A}})^{T} =\mathbf{\bar{B}^{-1}\bar{A}},\ \ (%
\mathbf{\bar{D}\bar{B}^{-1}})^{T}=\mathbf{\bar{D}\bar{B}^{-1}}
\label{ABCD-relation} \\
&&\mathbf{\bar{C}} -\mathbf{\bar{D}\bar{B}^{-1}\bar{A}}=-(\mathbf{\bar{B}}%
)^{T},
\end{eqnarray}%
among the sub-matrices of Eq.(\ref{V}). These will be used in
derivations that follow.

In conventional optics, the cross-spectral density of a partially
coherent wave is usually described by the
Gaussian-Schell-model(GSM)\cite{optical coherence} expression
\begin{equation}\label{GSML}
W(\bm\rho _{1},\bm\rho _{2})=\sqrt{S(\bm\rho _{1})}\sqrt{S(\bm\rho _{2})}g(%
\bm\rho _{1},\bm\rho _{2}),
\end{equation}%
where $S(\bm\rho )$ represents the spectral density, and $g(\bm{\rho}_{1},%
\bm{\rho}_{2})$ represents the spectral degree of coherence, given
by
\begin{eqnarray}
S(\bm{\rho})&=&G_{0}\exp [-\frac{1}{2}\bm{\rho\ }^{T}(\bm\sigma
_{s}^{2})^{-1}\bm{\rho\ }], \\
g(\bm{\rho}_{1},\bm{\rho}_{2})&=&\exp [-\frac{1}{2}(\bm{\rho}_{1}-\bm{\rho}%
_{2})^{T}(\bm\sigma _{g}^{2})^{-1}(\bm{\rho}_{1}-\bm{\rho}_{2})],
\end{eqnarray}%
where
\begin{equation}
\bm{\rho_1}^{T}=(x_{1},y_{1}),\ \ \bm{\rho_2}^{T}=(x_{2},y_{2})
\end{equation}%
are position vectors of two arbitrary points in the transverse
plane. $G_{0}$ is a positive quantity. $\bm\sigma _{s}$ represents
the transverse spot
width and $\bm\sigma _{g}$ represents the transverse coherent width. $\bm%
\sigma _{s}$ and $\bm\sigma _{g}$ are $2\times 2$ matrices with
transpose symmetry.

This partial coherence theory in optics can be extended to describe
the PCMW according to the analogy between the Schr\"{o}dinger
equation (\ref {heisenberg}) of a matter wave and the paraxial wave
propagation equation of light
\begin{equation}\label{para}
\nabla _{T}^{2}A(\mathbf{q})+2ik\frac{\partial
A(\mathbf{q})}{\partial z}=0,
\end{equation}%
where $A(\mathbf{q})$ is the amplitude of the electric field as
expressed by $E(\mathbf{q},t)=A(\mathbf{q})\exp [i(kz-\omega t)]$,
and $k=2\pi /\lambda $ is the modulus of the wave vector. The
squared transverse Laplacian $\nabla _{T}^{2}$ is mathematically
analogous to the atomic kinetic energy $-(\hbar
^{2}/2m)\nabla ^{2}$. The difference between Eq.(\ref{heisenberg}) and Eq.(%
\ref{para}) is that Eq.(\ref{heisenberg}) has a time derivative
while the paraxial wave equation Eq.(\ref{para}) has a spatial
derivative along the propagation axis $z$. The GSM partially
coherent light as described by Eq.( \ref{GSML}) has two spatial
dimensions perpendicular to the propagation axis $z$, while the GSM
of the PCMW may have three spatial dimensions. The first-order
correlation function of a GSM matter wave can be expressed as
\begin{eqnarray}\label{GSMm}
\Gamma(\mathbf{q}_1,\mathbf{q}_2)&=G_0\exp\Big\{ -
\frac{1}{4}\left[\mathbf{q}_1^{T}(\bm\sigma_{s}^2)^{-1}\mathbf{q}_1+\mathbf{q}_2^{T}(\bm\sigma_{s}^2)^{-1}\mathbf{q}_2\right]\nonumber\\
&-\frac{1}{2}(\mathbf{q}_1-\mathbf{q}_2)^{T}(\bm\sigma_{g}^2)^{-1}(\mathbf{q}_1-\mathbf{q}_2)\Big\},
\end{eqnarray}
where both the atom number density and the coherence degree have
Gaussian distributions. The parameters $G_{0}$, $\bm\sigma _{s}$ and
$\bm\sigma _{g}$ are related to the temperature of the atom gas.

Equation (\ref{GSMm}) can be rewritten as
\begin{equation}\label{GSM}
\Gamma (\mathbf{r})=G_{0}\exp \left( \frac{im}{2\hbar }\mathbf{r}^{T}%
\mathbf{M}_{i}^{-1}\mathbf{r}\right) ,
\end{equation}%
with $\mathbf{r}^{T}=(\mathbf{q}_{1},\mathbf{q}_{2})$. The tensor
\begin{eqnarray}\label{Mi}
\mathbf{M}_i^{-1}=\left[
\begin{array}{cc}
\frac{i\hbar}{2m}(\bm\sigma_{s}^{2})^{-1}
+\frac{i\hbar}{m}(\bm\sigma_{g}^{2})^{-1}& -\frac{i\hbar}{m}(\bm\sigma_g^{2})^{-1}\\
-\frac{i\hbar}{m}(\bm\sigma_g ^{2})^{-1} &
\frac{i\hbar}{2m}(\bm\sigma_{s}^{2})^{-1}
+\frac{i\hbar}{m}(\bm\sigma_{g}^{2})^{-1}
\end{array}
\right]
\end{eqnarray}
is a $6\times 6$ symmetric matrix, which may be called the
complex curvature tensor of the PCMW; where $\bm\sigma _{s}$ and
$\bm\sigma _{g}$ are $3\times 3$ matrices with transpose symmetry,
given by
\begin{equation}
\bm\sigma _{s}=\left[
\begin{array}{ccc}
\sigma _{sxx} & \sigma _{sxy} & \sigma _{sxz} \\
\sigma _{syx} & \sigma _{syy} & \sigma _{syz} \\
\sigma _{szx} & \sigma _{szy} & \sigma _{szz}%
\end{array}%
\right] ,
\end{equation}%
\begin{equation}
\bm\sigma _{g}=\left[
\begin{array}{ccc}
\sigma _{gxx} & \sigma _{gxy} & \sigma _{gxz} \\
\sigma _{gyx} & \sigma _{gyy} & \sigma _{gyz} \\
\sigma _{gzx} & \sigma _{gzy} & \sigma _{gzz}%
\end{array}%
\right] .
\end{equation}%
The tensor expression can be used to describe a general matter wave
system including the asymmetric and anisotropic systems, and is very
convenient for its compactness. The matrix $\bm\sigma _{s}$,
representing widths of the matter wave in 3D space, characterizes
the spatial size of the matter wave. The matrix $\bm\sigma _{g}$
describes the coherent length of the matter wave. When $\bm\sigma
_{g}$ decreases to zero, it corresponds to a completely incoherent
matter wave. When $\bm\sigma _{g}$ increases to infinity, the matter
wave can be treated as a completely coherent matter wave.

Substituting Eq.(\ref{deltaS}) and Eq.(\ref{GSM}) into
Eq.(\ref{Gamma}), we get
\begin{eqnarray}
\Gamma (\mathbf{r},t)&=& -\left( \frac{m}{ih}\right)
^{3}\frac{G_{0}}{|\det
\mathbf{B}|}   \label{Gammaout}\nonumber \\
&& \cdot \int \exp \left[ \frac{im}{2\hbar }\left( Q+2\dot{\bm\xi }\cdot (%
\mathbf{q}_{1}-\mathbf{q}_{2})\right) \right] d\mathbf{r}_{0},
\end{eqnarray}%
where
\begin{eqnarray}\label{L}
Q&=&(\mathbf{r}-\bar{\bm\xi})^{T}(\bar{\mathbf{C}}+\bar{\mathbf{D}}\mathbf{M}_i^{-1})(\bar{\mathbf{A}}+\bar{\mathbf{B}}\mathbf{M}_i^{-1})^{-1}(\mathbf{r}-\bar{\bm\xi})\nonumber\\
&&+\Big|(\mathbf{\bar{B}^{-1}\bar{A}}+\mathbf{M}_i^{-1})^{1/2}\mathbf{r}_0
-(\mathbf{\bar{B}^{-1}\bar{A}}+\mathbf{M}_i^{-1})^{-1/2}\mathbf{\bar{B}}^{-1}(\mathbf{r}-\bar{\bm\xi})
\Big|^2 .\end{eqnarray}
In the derivation of Eq.(\ref{L}), the transpose-symmetric
property of $\mathbf{M}_{i}^{-1}$ and Eq.(\ref{ABCD-relation}) have
been used. Integrating Eq.(\ref{Gammaout}), we have
\begin{eqnarray}
\Gamma (\mathbf{r})&=& G_{0}[\det (\bar{\mathbf{A}}+\bar{\mathbf{B}}\mathbf{M}%
_{i}^{-1})]^{-1/2}\exp \left[ im\dot{\bm\xi }\cdot (\mathbf{q}_{2}-\mathbf{q%
}_{1})/\hbar \right]   \label{ABCDlaw}\nonumber \\
&& \times \exp \left[ \frac{im}{2\hbar }(\mathbf{r}-\bar{\bm\xi })^{T}%
\mathbf{M}_{f}^{-1}(\mathbf{r}-\bar{\bm\xi })\right] ,
\end{eqnarray}%
where $\mathbf{M}_{i}^{-1}$ and $\mathbf{M}_{f}^{-1}$ denote the
complex curvature tensors of the PCMW at the initial and the final
time, respectively. They satisfy the condition
\begin{equation}\label{M}
\mathbf{M}_{f}^{-1}=(\bar{\mathbf{C}}+\bar{\mathbf{D}}\mathbf{M}_{i}^{-1})(%
\bar{\mathbf{A}}+\bar{\mathbf{B}}\mathbf{M}_{i}^{-1})^{-1}.
\end{equation}%
In the derivation of Eq.(\ref{ABCDlaw}), the integral formula
$\int_{-\infty }^{\infty }\exp [-ax^{2}]dx=\sqrt{\pi /a}$ has been
used. Equation (\ref{M}) may be called the tensor ABCD law for the
PCMW.

\section{Evolution of the PCMW in a gravitational field}

\label{example} To illustrate the usage of the formulas derived
above, we are going to calculate the atom density distribution of
the PCMW in a transverse plane in a gravitational field, which is
the diagonal element of the first-order correlation function $\Gamma
(\mathbf{q},\mathbf{q})$. The object we consider is a matter wave of
cold $Rb^{87}$ atoms. It propagates along the $z$ direction, and the
gravitational acceleration $\mathbf{g}$ with a linear gradient
$\gamma $ goes in the same direction. As is the case in
Ref.\cite{coq06}, the cold atom gas is prepared in a magnetic trap
with
trapping frequencies of $\omega _{x}=\omega _{z}=2\pi \times 330$ Hz and $%
\omega _{y}=2\pi \times 8$ Hz. The initial temperature of the
$Rb^{87}$ atom gas is assumed to be the recoil temperature by laser
cooling \cite{cooling}: $T_{rc}=362$ nk.

The propagation matrices $\mathbf{A,B,C}$ and $\mathbf{D}$ are
\begin{eqnarray}
\mathbf{A}& =&\left(
\begin{array}{ccc}
1 & 0 & 0 \\
0 & 1 & 0 \\
0 & 0 & A_{z}%
\end{array}%
\right) ,\ \mathbf{B}=\left(
\begin{array}{ccc}
t-t_{0} & 0 & 0 \\
0 & t-t_{0} & 0 \\
0 & 0 & B_{z}%
\end{array}%
\right) ,   \label{ABCDmatrix} \\
\mathbf{C}& =&\left(
\begin{array}{ccc}
0 & 0 & 0 \\
0 & 0 & 0 \\
0 & 0 & C_{z}%
\end{array}%
\right) ,\ \mathbf{D}=\left(
\begin{array}{ccc}
1 & 0 & 0 \\
0 & 1 & 0 \\
0 & 0 & D_{z}%
\end{array}%
\right) ,
\end{eqnarray}%
where $A_{z}=D_{z}=\cosh [\sqrt{\gamma }(t-t_{0})]$, $B_{z}=\sinh [\sqrt{%
\gamma }(t-t_{0})]/\sqrt{\gamma }$, and $C_{z}=\gamma B_{z}$. The
gravitational displacement vector is $\bm\xi ^{T}=\left( 0,0,\xi
_{z}\right)
$, and $\xi _{z}=(g/\gamma )[1-\cosh (\sqrt{\gamma }(t-t_{0}))]$ \cite{borde}%
.

Assuming the ultracold atoms are initially in a magnetic quadratic
trap, the widths of the atom gas in the $x,y,z$ directions
are\cite{BEC}
\begin{equation}
\sigma _{sj}= \left(\frac{k_{B}T}{m \omega _{j}^2}\right)^{1/2} ,
\end{equation}
where $\omega _{j}\ (j=x,y,z)$ is the harmonic oscillator frequency
in each dimension. Substituting the initial temperature $T_{rc}$ and
the initial trapping frequencies $\omega _{j}$ into the expression
of $\sigma_{sj}$, the widths of the matter wave source can be
obtained: $\sigma _{sxx}=2.84\mu \mbox{m},\ \sigma _{syy}=117.25\mu
\mbox{m}$. As we have assumed that the ultracold atom gas propagates
continuously along the $z$
direction, the width along the $z$ axis is infinite: $\sigma _{szz}=\infty $%
. Then we can write out the width tensor of the matter wave source
as
\begin{equation}
(\bm\sigma _{s}^{2})^{-1}=\left(
\begin{array}{ccc}
\sigma_{sx}^{-2} & 0 & 0 \\
0 & \sigma_{sy}^{-2} & 0 \\
0 & 0 & 0%
\end{array}%
\right) \mbox{mm}^{-2}.
\end{equation}%
We assume the coherent length $l_{c}$ to be $100\lambda_{T}$, which
is between the thermal de Borgile wavelength $\lambda_{T}=\sqrt{2\pi
\hbar^2/(mK_{B}T)}$ and the infinity. The coherent length matrix is
\begin{equation}
(\bm\sigma _{g}^{2})^{-1}=\left(
\begin{array}{ccc}
l_{c}^{-2} & 0 & 0 \\
0 & l_{c}^{-2} & 0 \\
0 & 0 & 0%
\end{array}%
\right) \mbox{mm}^{-2}.
\end{equation}%
Substituting the width matrix $(\bm\sigma _{s}^{2})^{-1}$ and the
coherent length matrix $(\bm\sigma _{g}^{2})^{-1}$ into
Eq.(\ref{Mi}), we can
calculate the final complex curvature tensor $\mathbf{M}_{f}^{-1}$ from Eq.(%
\ref{M}) and get the atom density distributions from
Eq.(\ref{ABCDlaw}). The expression of $\mathbf{A,B,C}$ and
$\mathbf{D}$ in Eq.(\ref{ABCDmatrix}) shows that the expansion of
the matter wave is dependent on time, and the transverse density
distributions at different temporal points reveal the evolution of
the matter wave as it propagates.

\begin{figure*}
\includegraphics[width=6cm]{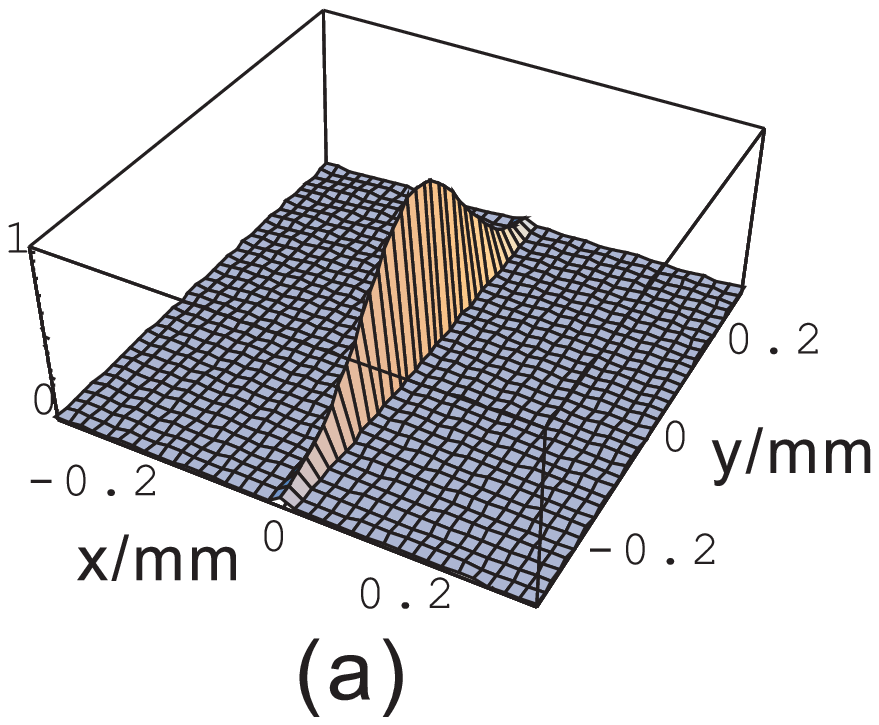}
\includegraphics[width=6cm]{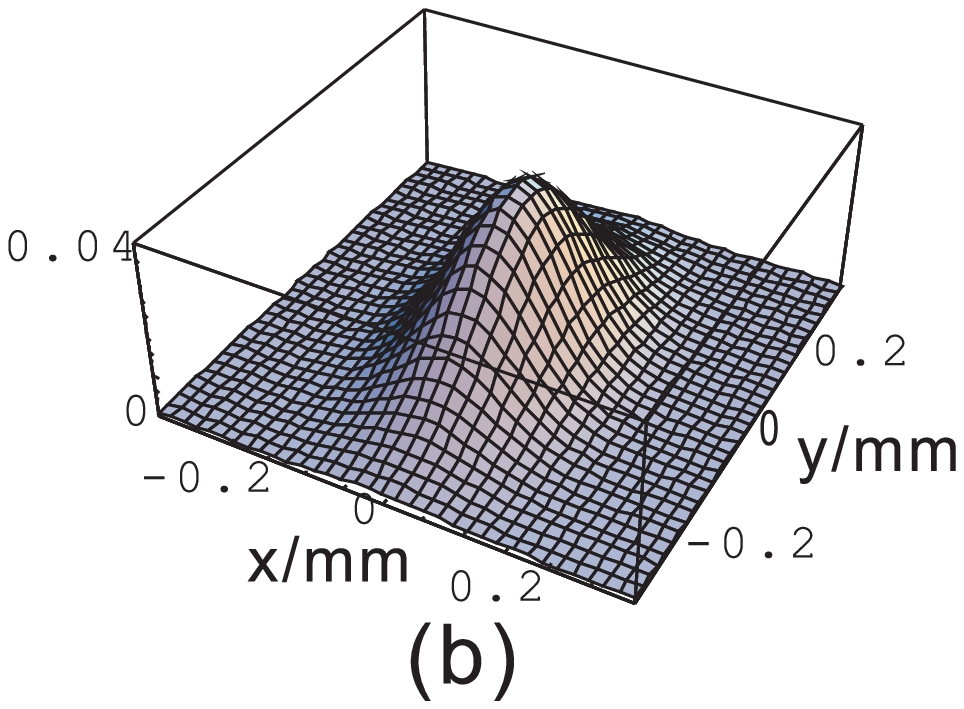}
\includegraphics[width=6cm]{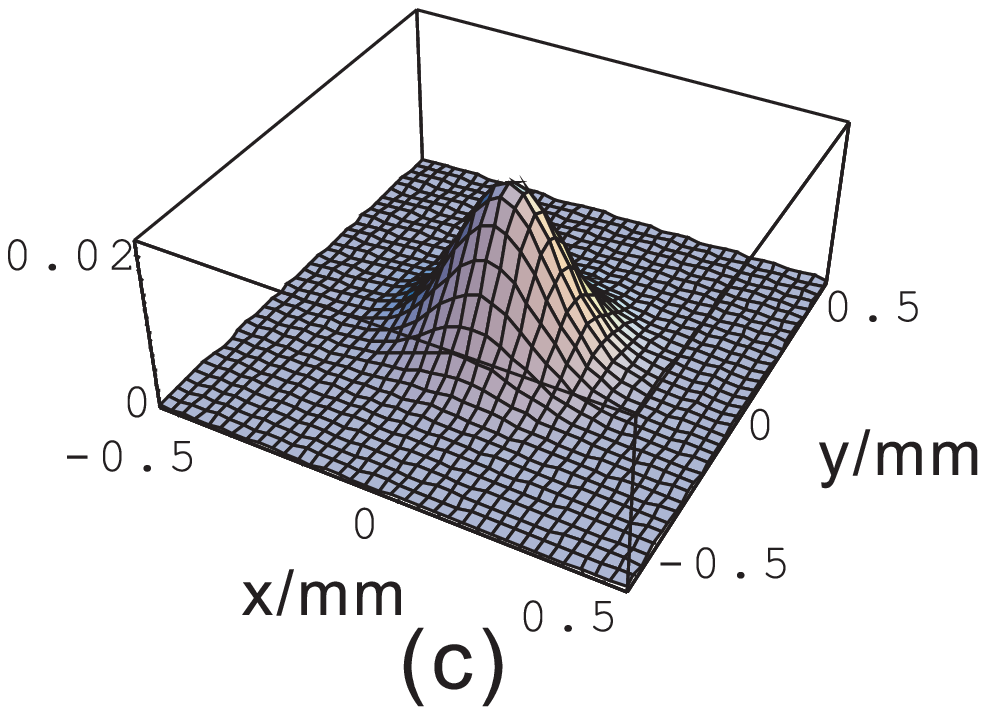}
\includegraphics[width=6cm]{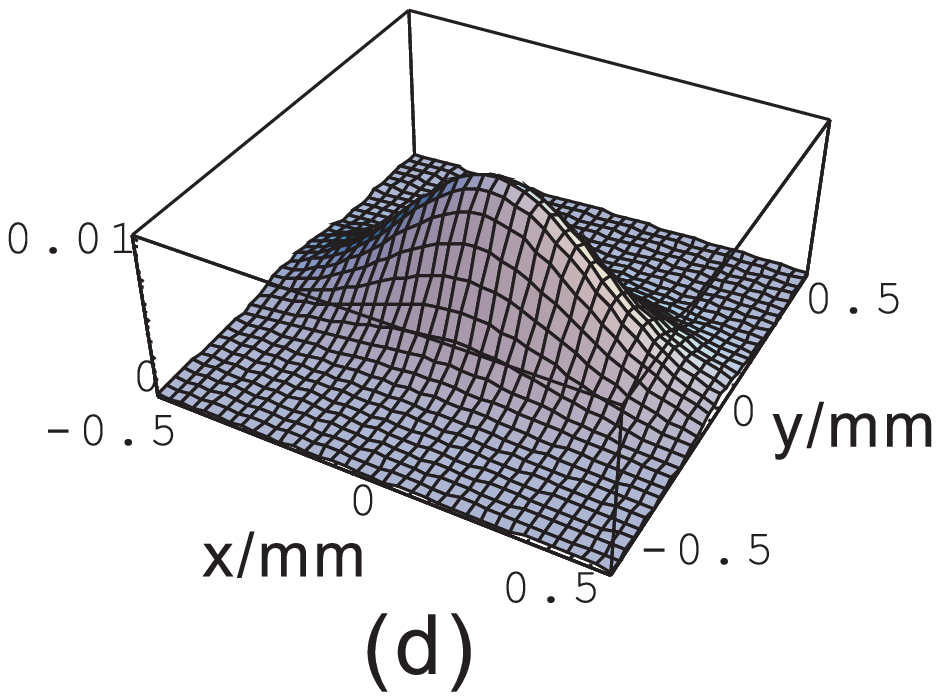}
\caption{The transverse relative atom density distribution for an
anisotropic GSM PCMW at different propagating times $t$. (a) $t=0s$; (b) $%
t=0.5s$; (c) $t=1s$; (d) $t=2s$.} \label{figure}
\end{figure*}

Figure \ref{figure} shows the evolution of the PCMW, with the
parameters given in the caption. The initial atom density profile,
which is shown in Fig.\ref{figure}(a), is elliptical due to the
initial inhomogeneous confinement of the cold atom gas on the x-y
plane; that is, the confinement in the x-direction is stronger,
which causes a larger speed of expansion of the matter wave in this
direction when the confinement is removed. During evolution without
confinement, the atom density profile expands gradually (
Fig\ref{figure}(b)), and changes to a circle (Fig.\ref{figure}(c)).
Finally the atom density profile becomes an ellipse again with the
major axis
perpendicular to that of the initial ellipse, as shown in Fig.\ref{figure}%
(d). This is because the degree of expansion in the x-direction
exceeds that in the y-direction. This property is similar to that of
light beams.
\section{Conclusion}
\label{conclusion} In this paper, a theoretical description of a
partially coherent matter wave (PCMW) based on the analogy between a
matter wave and an optical wave is presented. The propagation and
evolution formula of the Gaussian-Schell-Model PCMW is provided, and
a generalized tensor ABCD law is derived. As an example, we analyzed
the atom density profile in a transverse plane of a $Rb^{87}$ cold
atom beam. The results show that the tensor ABCD law is a very
convenient method for treating the propagation and transformation of
partially coherent cold atom beams. Most importantly, the previous
results for partially coherent light can be converted into a PCMW.
Our method can be applied to analyze the propagation and evolution
properties of ultracold atom beams in various potentials, such as an
optical potential and a magnetic potential. This method can also be
extended to pulsed matter wave propagation, and to include the
interactions among atoms and the internal state of atoms with some
modifications. These will be the topics of further investigations.

\appendix{\textbf{Acknowledgments}}

We wish to acknowledge support from the Natural Scientific
Foundation of China (grant no. 10574110) and the National Key
Project for Basic Research (grant no. 2006CB921403).

\end{document}